\documentclass[useAMS,usenatbib]{mn2e}
\usepackage{graphicx}

\newcommand{\xmm}{{\it XMM~\/}}
\newcommand{\xmmn}{{\it XMM-Newton~\/}}

\def\H0{{\rm ~km~s^{-1}~Mpc^{-1}}}

\def\la{\mathrel{\hbox{\rlap{\hbox{\lower4pt\hbox{$\sim$}}}{\raise2pt\hbox{$<$}}}}}
\def\ga{\mathrel{\hbox{\rlap{\hbox{\lower4pt\hbox{$\sim$}}}{\raise2pt\hbox{$>$}}}}}
\def\d25{D$_{25}$}

\def\deg{\hbox{$^\circ$}}

\title[The X-ray spectral variability of MCG--6-30-15] {Exploring the
  X-ray spectral variability of MCG--6-30-15 with \emph{XMM-Newton}}
\author[J.~Larsson, A.C.~Fabian, G.~Miniutti and R. R.~Ross]{
  J.~Larsson$^1$, A. C.~Fabian$^1$, G.~Miniutti$^1$ and R. R.~Ross$^2$\\
  $^1$Institute of Astronomy, Madingley Road, Cambridge CB3 0HA\\
  $^2$Physics Department, College of the Holy Cross, Worcester, MA
  01610, USA } \date{} 

\pagerange{\pageref{firstpage}--\pageref{lastpage}}
\pubyear{2007}

\begin{document}

\date{Accepted 2006 December 19}

\maketitle

\label{firstpage}

\begin{abstract}
  We present a study of the spectral variability of the Seyfert I
  galaxy MCG--6-30-15 based on the two long \xmmn observations from
  2000 and 2001.  The X--ray spectrum and variability properties of
  the 2001 data have previously been well described with a
  two-component model consisting of a variable power law and a much
  less variable reflection component, containing a broad relativistic
  iron line from the accretion disc around a rapidly rotating Kerr
  black hole. The lack of variability of the reflection component has
  been interpreted as an effect of strong gravitational light bending
  very close to the central black hole. Using an improved reflection
  model, we fit the two-component model to time-resolved spectra of
  both observations.  Assuming that the photon index of the power law
  is constant we reconfirm the old result and show that this does not
  depend on the time-scale of the analysis.
\end{abstract}


\section{Introduction}

The Seyfert I galaxy MCG--6-30-15 is a S0-type galaxy located at a
distance of 37 Mpc ($z=0.00775$) in the constellation of Centaurus.
Because of a very prominent broad, relativistic iron line in its
X--ray spectrum (first observed by \emph{ASCA}, Tanaka et al. 1995), this
object has come to play an important role in studies of accretion on
to black holes.

The spectrum of MCG--6-30-15 can be decomposed into two main
components, a simple power law component and a
relativistically-blurred reflection component. The reflection
component arises as the power law continuum irradiates the disc, and
its appearance is affected by relativistic effects close to the
central black hole.  The most prominent feature in the resulting 3--10
keV reflection spectrum is the broad, skewed Fe K $\alpha$ line
extending from about 3 to 7 keV and peaking at 6.4 keV. The observed
line profile and the shape of the overall reflection spectrum in this
source represents a fundamental tool for investigating the properties
of the accretion flow and the space time geometry close to the black
hole. In particular, the red wing of the iron line can be used to
determine the inner radius of the accretion disc and thereby the spin
of the black hole (see Brenneman \& Reynolds (2006) for a recent
determination of the black hole spin of MCG--6-30-15).

If the iron line is due to reflection one would, in a simple picture,
expect that its strength (and the strength of the whole reflection
component) follows the intensity of the power law continuum, which
irradiates the disc and gives rise to the reflection component.
Analysis of the variability of the two components has however shown
that the reflection component remains nearly constant while the power
law shows large variations (Shih et al 2002; Fabian et al 2002; Fabian
\& Vaughan 2003; Taylor, Uttley \& McHardy 2003, Papadakis, Kazanas \&
Akylas 2005). The lack of a correlation between the two components and
the small variability of the reflection component can be explained by
effects of strong gravitational light bending close to the central
black hole (Fabian $\&$ Vaughan 2003; Miniutti et al.  2003; Miniutti
$\&$ Fabian 2004), or by an inhomogeneous accretion disc (Merloni et
al. 2006). In both cases, the reflection spectrum has to be emitted
from the innermost regions of the accretion disc where relativistic
effects dominate.

In this Paper, we use the two long \xmmn observations from 2000 and
2001 to re--analyse the spectral variability of MCG--6-30-15 by fitting
time-resolved spectra. This has previously been done for 10 ks spectra
of the 2001 observation (Vaughan $\&$ Fabian 2003), showing that the
power law varies with an almost constant slope while the reflection
component varies by less than 25 per cent.

Since the time of this analysis, a new, more complete reflection model
has been made available (Ross $\&$ Fabian, 2005). The new model
includes all energetically important ionisation states and transitions
in the accretion disc and should therefore be a better approximation
of the real conditions. The most notable difference in the 3--10 keV
reflection spectrum is that the Fe K edge is much smoother than before
due to the presence of a larger range of ionisation states of iron.
It is important to test if (or how) the results change when using this
refined reflection model and this is an important reason for repeating
the analysis by Vaughan $\&$ Fabian (2003).

We also extend the original analysis by including the observation from
2000 which catches MCG--6-30-15 in a low flux state. This observation
makes up nearly half the data below the mean flux of both observations
and thereby adds valuable information about the source, which seems to
behave differently at low fluxes.  For the low flux (2000) observation
Ponti et al. (2004) found possible iron line variations and Reynolds
et al. (2004) reported on a correlation between the iron line
intensity (representative of the whole reflection component) and the
continuum flux. A correlation between the two components at low fluxes
and no correlation at higher fluxes is consistent with the predictions
of the light bending model.  By including the data from 2000 we aim to
confirm this result by performing the same analysis on both
observations.

It is also of interest to investigate whether the variability
shows any dependence on the time-scale. We therefore perform the
fitting on 3 ks and 30 ks spectra as well as on  10 ks spectra.


\section{observations and data reduction}

MCG--6-30-15 has been observed by \xmmn during revolution 108 (2000
July 11-12) and revolutions 301-303 (2001 July 31-August 5). The
observations are described in detail in Wilms et al. (2001) and
Vaughan $\&$ Fabian (2004). For this work the data was reprocessed
entirely using the \xmmn Science Analysis System ({\scriptsize SAS
  V}6.5.0), following the procedure described in Vaughan $\&$ Fabian
(2004), but only using the data from the EPIC-pn camera.  The total
amount of good exposure time selected was 84 ks and 318 ks for the
2000 and 2001 observations.
 
\section{spectral analysis}

The goal of the following analysis is to investigate the relative
variability of the power law and reflection components. This is done
by fitting the same model to time-resolved spectra from both
observations, leaving the relevant parameters free. In order to
determine the exact model to be used when fitting the time-resolved
spectra, we first fit the time-averaged spectra from all 4 orbits.

All spectra were fitted using {\scriptsize XSPEC V}11.3.2 (Arnaud
1996) and all fit parameters are quoted at the rest frame of the
source.  The analysis is limited to the 3--10 keV range in order to
avoid problems due to the complex warm absorption at soft X-ray
energies.

\subsection{Time-averaged spectra}\label{timeav}

\begin{figure}
\rotatebox{270}{\resizebox{!}{8 cm}{\includegraphics{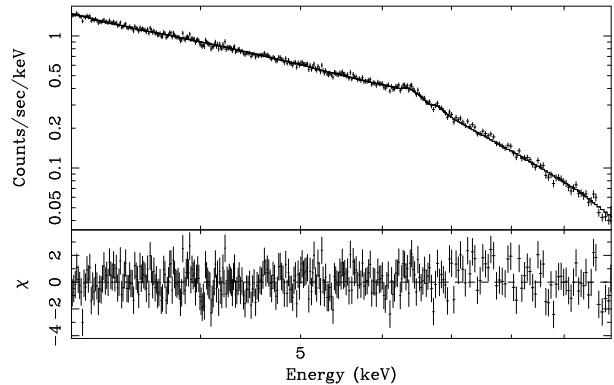}}}
\caption{\xmm EPIC pn spectrum of the second orbit in 2001, fitted with
  the model described in section \ref{timeav}}\label{spectrum}
\end{figure}
\begin{figure}
\rotatebox{270}{\resizebox{!}{8 cm}{\includegraphics{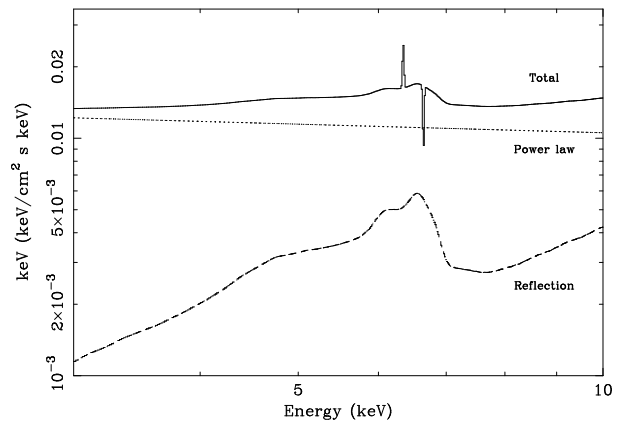}}}
\caption{The spectral model (solid) together with its main components;
  the power law (dotted) and the relativistically-blurred reflection
  component (dashed). The broad iron line is clearly
  visible in the reflection component.}\label{model}
\end{figure}

Time-averaged spectra of all four orbits were fitted simultaneously
with a spectral model consisting of a power law, a
relativistically-blurred reflection component, a narrow emission line
at 6.4 keV (from distant neutral iron), and a narrow absorption line
at 6.7 keV (from Fe {\small XXV}) as in Vaughan $\&$ Fabian (2004) and
Young et al. (2005).

To calculate the reflection spectrum, the model {\scriptsize REFLIONX}
(a high-resolution version of {\scriptsize REFLION}, Ross $\&$ Fabian
2005) was used. The model calculates the emission from a photoionised
accretion disc in thermal and ionisation equilibrium. The input
parameters are the photon index of the incident power law ($\Gamma$),
the iron abundance, the ionisation parameter of the disc ($\xi = 4\pi
F_{\rm x}/n_{\rm H}$), and the normalisation. To account for Doppler
and gravitational effects, the reflection spectrum was convolved with
the relativistic kernel {\scriptsize KDBLUR}2 which is a modified
version of the {\scriptsize LAOR} code (Laor 1991). The relativistic
kernel model assumes an emissivity profile of the form $\epsilon
=r^{-q_{\rm{in}}}$ inside a breaking radius, $r_{\mathrm{br}}$, and
$\epsilon =r^{-q_{\rm{out}}}$ outside. In addition to the parameters
describing the emissivity profile, the model takes as input parameters
the inner and outer radii of the disc ($r_{\mathrm{in}}$,
$r_{\mathrm{out}}$) and the inclination ($i$).

The time-averaged spectra were fitted simultaneously with only the
normalisations for the power law and reflection components allowed to
vary between the orbits.  The best-fitting parameters ($\chi^2 =
4384/4493$ dof) correspond to a weakly ionised disc
($\xi=43.7^{+28.1}_{-1.3} \ \mathrm{erg\ cm\ s^{-1}}$) with an iron
abundance of $3.4^{+0.2}_{-0.1}$ times solar, extending from
$r_{\mathrm{in}}=1.9^{+0.1}_{-0.1}\ r_{\rm g}$ to
$r_{\mathrm{out}}=400 \ r_{\rm g}$ (the maximum allowed value), where
$r_{\rm g}=GM/c^2$ is the gravitational radius. The emissivity profile
of the disc is given by $q_{\mathrm{in}}=5.3^{+0.1}_{-0.2}$,
$q_{\mathrm{out}}=2.3^{+0.1}_{-0.1}$, and
$r_{\mathrm{br}}=6.7^{+0.1}_{-0.1}\ r_{\rm g}$. The inclination angle
is $i=39.1^{+0.8}_{-0.6}$ and the photon index was found to be
$\Gamma=2.19^{+0.04}_{-0.04}$. When fitted separately, this model gave
equally good fits to all the time-averaged spectra, and there was no
indication of the disc parameters changing between the observations.
The spectrum of the second orbit in 2001 fitted with the model is
shown in Fig.  \ref{spectrum} and the model together with its main
components is shown in Fig. \ref{model}.
\begin{figure}
\rotatebox{270}{\resizebox{!}{8 cm}{\includegraphics{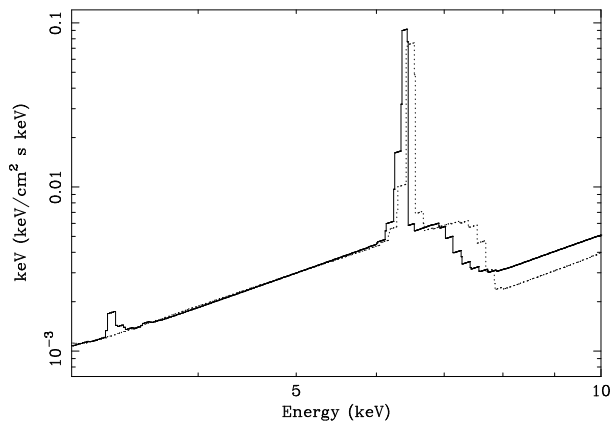}}}
\caption{Comparison of new (solid line) and old (dotted line)
  reflection models for $\xi = 40\ \mathrm{erg\ cm\ s^{-1}}$. In the
  new model the Fe K edge is smeared due to the presence of a larger
  range of ionisation states. Note also the sulphur line at low
  energies which is not present in the old model.}\label{comp}
\end{figure}

Figure 3 shows a comparison between the old and new reflection models
for $\xi = 40\ \mathrm{erg\ cm\ s^{-1}}$ and an iron abundance of 3.
The two models differ significantly around the iron line and edge
between 6 keV and 8 keV. This is due to the presence of lower
ionisation states of iron in the new model.  For the given parameters
the dominant ion of iron in the new model varies from Fe VI (the least
ionised species treated) to Fe XVIII with decreasing Thomson depth,
and as a consequence the iron K edge runs from 7.2 keV to 7.8 keV. In
the old model the least ionised species is Fe XVI and the edge runs
from 7.6 keV to 7.8 keV.  Another noticeable difference in the 3--10
keV range is the sulphur line around 3.3 keV which is not present in
the old model. As a consequence of the larger number of species
included, the new model is also less ionised than the old one for a
given value of $\xi$. In terms of best fitting parameters these
important differences mainly manifest themselves in that the new model
favours a higher inclination ($40\deg$ compared to previous results of
around $30\deg$), due to the change of the blue end of the line seen
in Fig.  \ref{comp}.


\subsection{Spectral variability}

\begin{figure}
\resizebox{8 cm}{!}{\includegraphics{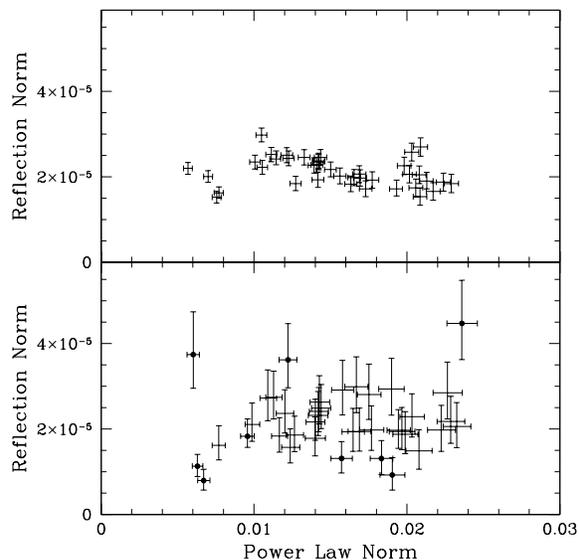}}
\caption{Results from fits to 10 ks spectra. Top panel: the reflection
  normalisation plotted against the power law normalisation for
  $\Gamma$ frozen at 2.15. Bottom panel: the reflection normalisation
  plotted against the power law normalisation for the case when
  $\Gamma$ is left as a free parameter. Black dots indicate the 9 over
  39 spectra for which a free $\Gamma$ value significantly improves
  the fit ($>$ 95 per cent level in an F-test). Note that these
  spectra tend to be the outliers.}\label{plc}
\end{figure}

The observations were split into 39 x 10 ks intervals (8 intervals
from the 2000 observation and 31 from the 2001 observation), and
spectra were extracted in each interval.  All the 10 ks spectra were
then fitted with the model described in the previous section, with all
parameters apart from the the power law and reflection normalisations
frozen. (Experimentation showed that the 10 ks spectra favoured a
slightly lower $\Gamma$ than the time-averaged spectra, 2.15 rather
than 2.19, and $\Gamma$ was therefore frozen at this value). The
results are shown in the upper panel of Fig. \ref{plc}. The two
components do not show any correlation and the reflection component is
seen to be much less variable than the power law, in excellent
agreement with previous findings. Although the variability of the
reflection component is small it is not consistent with being
constant; a $\chi^2$ fit to a constant gives $\chi^2=145/38$ dof,
which implies an intrinsic rms variability of about 10 per cent.

In order to investigate the robustness of these results all the fits
were performed with each of the parameters describing the accretion
disc left free. There was no parameter which significantly altered the
result or showed any correlation with continuum flux or time,
justifying our approach of using the same disc parameters to fit all
the spectra.

We also investigated the effect of leaving the photon index $\Gamma$
free. The resulting values of the two normalisations are shown in the
lower panel of Fig. \ref{plc}. Spectra for which thawing $\Gamma$
significantly improves the fit ($>$ 95 per cent level in an F-test)
are shown as black dots (9 spectra over 39). Only one spectrum
requires a different $\Gamma$ at the three sigma confidence level and
for the majority of spectra the fits improved only slightly or not at
all. The variability of the reflection component is seen to be roughly
the same as that of the power law (excluding the extreme 10 per cent
of both components) and about twice as high as that found by Fabian
$\&$ Vaughan (2003).

\begin{figure}
\resizebox{8 cm}{!}{\includegraphics{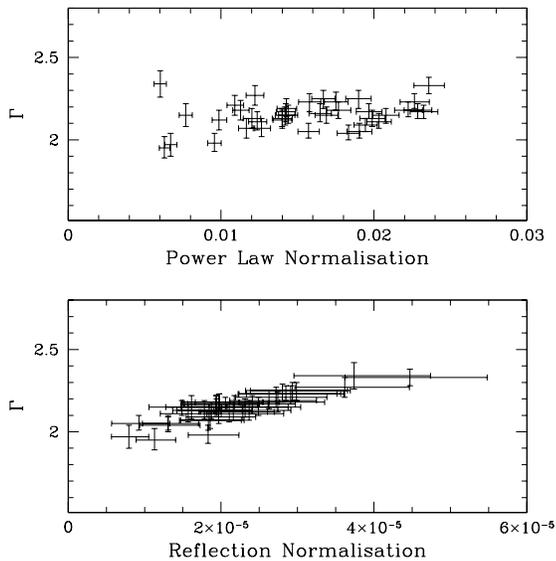}}
\caption{ Results from fits to 10 ks spectra. Top panel: the photon
  index $\Gamma$ of the power law as a function of the power law
  normalisation. Bottom panel: $\Gamma$ as a function of the
  reflection normalisation.}\label{gammas}
\end{figure}

The resulting values of $\Gamma$ as a function of the power law and
reflection normalisations are shown in Fig.  \ref{gammas}. $\Gamma$
lies around 2.1--2.2 in agreement with previous findings and we also
note that there is a correlation between the reflection normalisation
and $\Gamma$. Such a correlation has previously been reported by e.g.
Vaughan \& Edelson (2001), who found that the correlation was present
even when fitting spectra which were simulated with a constant
reflection fraction. We carried out a similar experiment and simulated
50 spectra with the same $\Gamma$ and reflection normalisation but
different power law normalisations. In order to mimic the real data as
closely as possible the reflection normalisation and $\Gamma$ were
fixed at their average values obtained from the 10 ks fits, and the
power law was given normalisations uniformly distributed over the
range seen in Fig. \ref{plc}.  Fitting of the simulated spectra did
indeed reveal similar results to those in the lower panel of Fig.
\ref{gammas}, showing that the relation is most likely an artifact of
the degeneracy of the two components. Since the reflection spectrum is
very hard in the 3--10~keV band, it is obvious that a steeper
continuum photon index requires more reflection than a flatter one.
This explains the observed spectral degeneracy between the continuum
slope and the reflection normalisation.
\begin{figure}
\rotatebox{270}{\resizebox{!}{8 cm}{\includegraphics{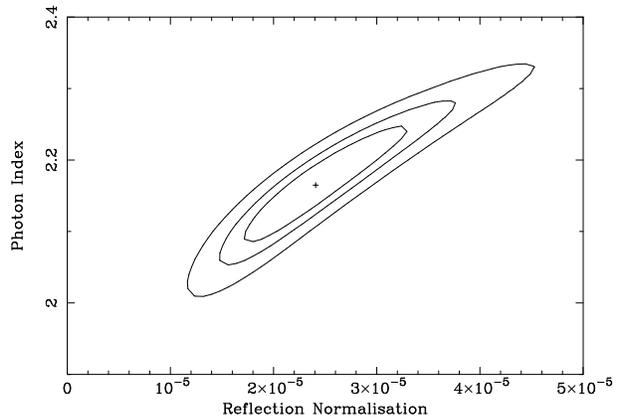}}}
\caption{Contours of 1, 2 and 3 sigma in the reflection normalisation,
  $\Gamma$--plane for one of the fitted 10 ks spectra. Note that the
  range is very similar to that found in the lower panel of Fig.
  \ref{gammas}}\label{cont}
\end{figure}

Fig. \ref{cont} shows contours of 1, 2 and 3 sigma in the reflection
normalisation, $\Gamma$--plane for one of the fitted spectra.  It is
clear from the figure that a spread in $\Gamma$ of about 0.2 will
result in a spread in reflection normalisation of the same order of
magnitude as that seen in the lower panel of Fig. \ref{plc}. Because
of this and of the spurious correlation between the reflection
normalisation and $\Gamma$ discussed above, it seems most likely that
the additional variability seen in the lower panel of Fig. \ref{plc}
is induced by the parameter degeneracy.

It should however be noted that the values of $\Gamma$ are
systematically lower at low fluxes (power law norms below 0.01) and
that freezing $\Gamma$ at 2.15 is not a good approximation in this
region. This is further confirmed by the fact that there is a
correlation between the two normalisations in this region when
$\Gamma$ is free.  A correlation at low fluxes and no correlation for
higher fluxes is consistent with previous findings (Reynolds et
al. 2004; Fabian $\&$ Vaughan, 2003) and with the predictions of the
light bending model (Miniutti \& Fabian 2004).

Leaving each of the disc parameters free in the fits with a free
$\Gamma$ does not affect the relationships between the parameters
shown in Figs. \ref{plc} and \ref{gammas} in any other way than by
introducing a larger spread.  The only disc parameter which changes
significantly when left free is the iron abundance, which goes all the
way up to 10 times solar. An iron abundance this high has also been
found by Brenneman \& Reynolds (2006) when using the same reflection
model. The fact that the new reflection model favours a higher iron
abundance is probably due to the difference in shape of the iron edge
seen in Fig. \ref{comp}.  Even though the iron abundance in
MCG--30-6-15 is known to be high, a value of 10 does not seem
plausible and we do indeed find it to be around 2--3 when including
high-energy data from \emph{ BeppoSAX}, \emph{RXTE} or \emph{Suzaku}
(Miniutti et al. 2006), as well as in the time-averaged fit described
in section \ref{timeav} and in all the fits where $\Gamma$ was frozen
(see also Lee et al.  2000).

In order to investigate whether the variability is dependent on the
time-scale of the analysis the same fits were also performed on 30 ks
and 3 ks spectra. These spectra give fits which are fully consistent
with the results from the 10 ks analysis, although the 3 ks fits have
much larger error bars due to the limited statistics.  By making a
finer grid in the reflection model we also established that the
results are not dependent on the resolution of the model.


\section{Conclusions and Discussion}

We have re--analysed the spectral variability of MCG--6-30-15 in terms
of the two-component model (power law + reflection from the accretion
disc) by fitting the model to time-resolved spectra taken from the two
long \xmmn observations of the source for a total of $\sim$390~ks. We
used a new reflection model ({\scriptsize REFLIONX}) and performed the
analysis on 3 ks, 10 ks and 30 ks spectra. The results were the same
for all three time-scales.  When only the normalisations of the two
components are allowed to vary the results are consistent with
previous findings in showing the reflection component to be much less
variable than the power law.

When the photon index $\Gamma$ of the power law is left as a free
parameter we however find that the two components vary by the same
amount, even though the spread in $\Gamma$ is very small for most
fluxes.  Since we also find an artificial correlation between the
reflection normalisation and $\Gamma$, it seems likely that this
variability is due to the obvious degeneracy between the continuum
slope and the amount of reflection rather than being real. The
artificial correlation strongly suggests that it is not possible to
simultaneously measure $\Gamma$ and the amount of reflection by direct
spectral fitting, and we therefore have to consider the results of
other, model-independent tools to assess the variability of the two
components.

\begin{figure}
\rotatebox{270}{\resizebox{!}{8 cm}{\includegraphics{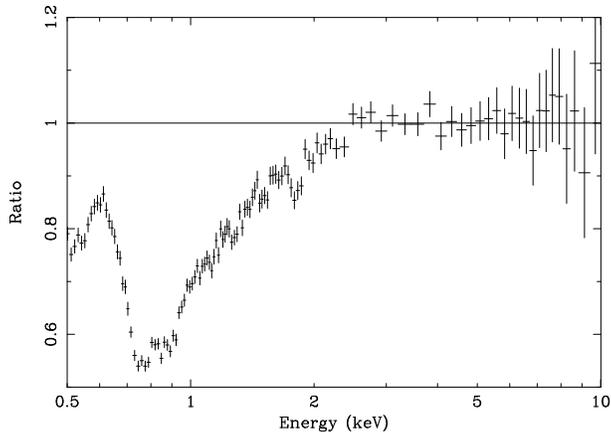}}}
\caption{Difference spectrum produced by subtracting a low-state
  spectrum from a high-state spectrum. The difference spectrum is
  shown as a ratio to a power law with $\Gamma=2.2$, modified by
  Galactic absorption and fitted over the 3--10 keV range.}\label{diff}
\end{figure}

One such tool is the difference spectrum between a high-flux and a
low-flux spectrum (e.g. Fabian \& Vaughan 2003). Under the assumption
that the total spectrum can be described by one constant and one
variable component, subtracting a low-state from a high-state spectrum
should remove the contribution of the constant component, leaving only
the variable component, modified by absorption. A difference spectrum
of MCG--6-30-15 based on both the 2000 and 2001 data is shown in Fig.
\ref{diff} as a ratio to a power law (and Galactic absorption of $4.06
\times 10^{20} \rm{cm}^{-2}$). The high and low states were
constructing by summing the third of the 10 ks spectra with the
highest and lowest fluxes respectively.  The power law was fitted over
the 3--10 keV energy range and has a photon index of 2.2 fully
consistent with the time--averaged value. The spectrum above 3 keV is
clearly well described by the power law and extrapolating to lower
energies reveal signs of strong, partially-ionised absorption. This
result is very similar to the difference spectrum presented by Vaughan
\& Fabian (2004), which was based only on the 2001 observation. The
lack of the Fe line in the difference spectrum demonstrates that the
reflection component is approximately constant and has been subtracted
away, confirming our results in a model--independent way.  We also
stress that the difference spectrum above 3~keV is extremely well
described by the power law, which rules out the presence of subtle
absorption in the hard band that may otherwise affect the broad
relativistic line parameters.
Other model-independent methods which have strongly suggested that the
reflection component varies very little are root-mean-squared spectra,
which measure the the variability as a function of energy, and
flux--flux plots (Taylor, Uttley \& McHardy 2003), which show the
relationship between fluxes in different energy bands (see e.g.
Vaughan \& Fabian 2004 for root-mean-squared spectra and flux--flux
plots of MCG--6-30-15).

Although none of these methods can rule out that there is a relatively
large variability in the reflection component (which is however
uncorrelated with the power law), they all indicate that the
reflection component varies much less than the power law and that the
photon index is nearly constant.  Because of this, and because most of
the fits do not improve significantly when $\Gamma$ is left as a free
parameter, we conclude that it is a safe assumption to freeze $\Gamma$
in the spectral fitting. Both the best-fitting model for the
time-averaged spectra and the difference spectrum give $\Gamma \approx
2.2$, and most of the 10 ks fits with $\Gamma$ free agree with this
value.  The only time it does not seem to be appropriate to freeze
$\Gamma$ at around 2.2 is at very low fluxes, where $\Gamma$ is seen
to be lower by $\Delta\Gamma\sim 0.2$.  With a low $\Gamma$ in this
region we do see the previously reported correlation between the two
normalisations. Under these assumptions we thus confirm the conclusion
that the strength of the reflection increases with increasing
continuum flux at low fluxes but then saturates when the source
reaches its normal state, in agreement with the predictions of the
light bending model, as reported also for e.g.  NCG~4051 (Ponti et al.
2006), 1H~0707--495 (Fabian et al. 2004), and the Galactic black hole
XTE~J1650--500 (Miniutti, Fabian \& Miller 2004; Rossi et al 2005).

\section*{Acknowledgements}
JL thanks Corpus Christi College, the Isaac Newton Trust and PPARC.
ACF and GM thank the Royal Society and the UK PPARC respectively for
support. RRR thanks the College of the Holy Cross for support. The
results presented here are based on observations obtained with
\emph{XMM-Newton}, an ESA science mission with instruments and
contributions directly funded by ESA Member States and NASA.

{}

\end{document}